\documentclass[amsmath,amssymb,twocolumn,prd,floatfix,showpacs, nofootinbib]{revtex4-1}
\usepackage{tabularx}  % tabular environment
\usepackage{bm, color} % math bold and /color
\usepackage{overpic,subfigure} % figures
\usepackage{multirow}
\usepackage{array}
\usepackage{dcolumn} % to deal with numbers with units and uncertainties
\usepackage[symbol]{footmisc}
\usepackage{epstopdf}
\usepackage{ulem}
\RequirePackage{xspace}

\newcommand{\gev}{\ensuremath{\mathrm{\,Ge\kern -0.1em V}}\xspace}
\newcommand{\mev}{\ensuremath{\mathrm{\,Me\kern -0.1em V}}\xspace}
\newcommand{\mevcc}{\ensuremath{{\mathrm{\,Me\kern -0.1em V\!/}c^2}}\xspace}

\newcommand{\jprBase}        {Phys.\ Rev.\xspace}
\newcommand{\jplBase}        {Phys.\ Lett.\xspace}
\newcommand{\plb}      [1]{\jplBase\ {\bf B~#1}}
\newcommand{\jprd}     [1]{\jprBase\ {\bf D~#1}}

\def\epem       {\ensuremath{e^+e^-}\xspace}

\def\kp         {\ensuremath{K^+}\xspace}
\def\km         {\ensuremath{K^-}\xspace}
\def\ks         {\ensuremath{K_{S}^{0}}\xspace}

\def\pip        {\ensuremath{\pi^+}\xspace}
\def\pim        {\ensuremath{\pi^-}\xspace}

\def\jpsi       {\ensuremath{{J\mskip -3mu/\mskip -2mu\psi\mskip 2mu}}\xspace}
\def\chic#1{\ensuremath{\chi_{c#1}}\xspace} % dbm

\def\fz#1       {\ensuremath{f_0({#1})}\xspace}

\begin{document}
%\linenumbers
%\title{\boldmath Observation of 4$K_S^0$ decay of $\chi_{cJ}$ meson}
\title{\boldmath Observation of $\chi_{cJ}\to4K_{S}^{0}$}

\author{
M.~Ablikim$^{1}$, M.~N.~Achasov$^{10,d}$, S. ~Ahmed$^{15}$, M.~Albrecht$^{4}$, M.~Alekseev$^{55A,55C}$, A.~Amoroso$^{55A,55C}$, F.~F.~An$^{1}$, Q.~An$^{52,42}$, Y.~Bai$^{41}$, O.~Bakina$^{27}$
% ...
\\
\vspace{0.2cm}
(BESIII Collaboration)\\
\vspace{0.2cm} {\it
$^{1}$ Institute of High Energy Physics, Beijing 100049, People's Republic of China\\
% ...
$^{4}$ Bochum Ruhr-University, D-44780 Bochum, Germany\\
% ...
$^{10}$ G.I. Budker Institute of Nuclear Physics SB RAS (BINP), Novosibirsk 630090, Russia\\
% ...
$^{15}$ Helmholtz Institute Mainz, Johann-Joachim-Becher-Weg 45, D-55099 Mainz, Germany\\
% ...
$^{27}$ Joint Institute for Nuclear Research, 141980 Dubna, Moscow region, Russia\\
% ...
$^{41}$ Southeast University, Nanjing 211100, People's Republic of China\\
$^{42}$ State Key Laboratory of Particle Detection and Electronics, Beijing 100049, Hefei 230026, People's Republic of China\\
% ...
$^{52}$ University of Science and Technology of China, Hefei 230026, People's Republic of China\\
% ...
$^{55}$ (A)University of Turin, I-10125, Turin, Italy; (B)University of Eastern Piedmont, I-15121, Alessandria, Italy; (C)INFN, I-10125, Turin, Italy\\
% ...
$^{d}$ Also at the Novosibirsk State University, Novosibirsk, 630090, Russia\\
% ...
}
\vspace{0.4cm}
}

\begin{abstract}
  %Using a data sample of $(448.1\pm2.9)\times 10^6~\psi'$ events collected with the BESIII detector at the BEPCII collider, we present measurements of branching fractions for the decays $\chi_{cJ}(J=0,1,2)\to$\fourks. The decay into the \fourks hadronic final state is
%  observed for the first time. We measure the branching fractions
%  ${B}(\chic{0}\to\fourks)=(5.73 \pm 0.34 \pm 0.40)\times10^{-4}$,
%  ${B}(\chic{1}\to\fourks)=(0.35 \pm 0.08 \pm 0.03)\times10^{-4}$, and
%  ${B}(\chic{2}\to\fourks)=(1.18 \pm 0.15 \pm 0.09)\times10^{-4}$, where
%  the uncertainties are statistical and systematical, respectively.
  By analyzing $(448.1\pm2.9)\times10^6$ $\psi(3686)$ events collected with the BESIII detector operating at the BEPCII collider, the decays of $\chi_{cJ} \to 4\ks$ ($J=0,1,2$) %$K_{S}^{0}\to\pi^+\pi^-$
  are observed for the first time with statistical significances of 26.5$\sigma$, 5.9$\sigma$ and 11.4$\sigma$, respectively. The product branching fractions of $\psi(3686)\to\gamma\chi_{cJ}$, $\chi_{cJ}\to4K_{S}^{0}$ are presented and the branching fractions of $\chi_{cJ}\to4K_{S}^{0}$ decays are determined to be
  $\mathcal{B}_{\chi_{c0}\to 4\ks}$=$(5.76\pm0.34\pm0.38)$$\times10^{-4}$,
  $\mathcal{B}_{\chi_{c1}\to 4\ks}$=$(0.35\pm0.09\pm0.03)$$\times10^{-4}$ and
  $\mathcal{B}_{\chi_{c2}\to 4\ks}$=$(1.14\pm0.15\pm0.08)$$\times10^{-4}$,
  where the first uncertainties are statistical and the second are systematic, respectively. %{\color{blue}The product branching fractions of $\psi(3686)\to\gamma\chi_{cJ}$, $\chi_{cJ}\to4K_{S}^{0}$ are also presented.}

\end{abstract}

% PACS, the Physics and Astronomy Classification Scheme.

\maketitle

\section{Introduction}
In the quark model, the \chic{J} ($J=0,1,2$) mesons are the $^3P_J$ charmonium
states. Since the $\chi_{cJ}$ mesons cannot be directly produced in $\epem$ collisions, according to parity conservation, their decays are experimentally and theoretically not studied as extensively as the vector charmonium states \jpsi and $\psi(3686)$. However, the $\chi_{cJ}$ mesons can
be produced in radiative decays of the $\psi(3686)$ with branching fractions of about 9\%, which provides a method to produce large $\chi_{cJ}$ samples in order to study $\chi_{cJ}$ decays.

Recent theoretical work indicates that the Color Octet Mechanism~(COM)~\cite{com} could have large contributions to the decays of the
$P$-wave charmonium states. However, many contradictions still exist between these theoretical calculations and experimental measurements. For instance, theoretical predictions of $\chi_{cJ}$ decays to baryon anti-baryon pairs based on the COM~\cite{ref::theroy1}\cite{ref::theroy2}\cite{ref::theroy3} were inconsistent with experimental measurements~\cite{pdg}. Thus more
precise experimental results are
mandatory to further understand \chic{J} decay dynamics.  Furthermore,
the \chic{0} and \chic{2} states are expected to decay via
two-gluon processes into light hadrons, giving access to the investigation of glueball dynamics. Thus,  comprehensive measurements of
exclusive hadronic decays of $\chi_{cJ}$ are valuable.

For the decay modes of $\chi_{cJ}\to4K$, the branching fractions of $\chi_{cJ}$ decays into 2($ K^{+}K^{-})$ and $ K^{+}K^{-}K_{S}^{0}K_{S}^{0}$ have been measured by Belle~\cite{ref::kkkk} and BES~\cite{ref::kkk0k0} with results summarized in Table~\ref{tab:bf}.
In this paper, by analyzing $(448.1\pm2.9)\times 10^6$ $\psi(3686)$ events~\cite{ref::psip-num-inc} collected with the BESIII detector~\cite{NIM10}, we present the first measurements of the branching fractions of $\chi_{cJ}$ decays to 4\ks.

\begin{table}[tb]
\begin{center}
\caption{World averages on branching fractions of $\chi_{cJ}$ decays to $2(\kp\km)$ and $\kp\km\ks\ks$ \cite{ref::kkkk, ref::kkk0k0, pdg}}.
\begin{tabular} {lr}
\hline\hline
Channel & Branching fraction ($\times 10^{-3}$)\\\hline
$\chic{0}\to 2(\kp\km)$ & $2.82\pm0.29$ \\
$\chic{1}\to 2(\kp\km)$ & $0.54\pm0.11$ \\
$\chic{2}\to 2(\kp\km)$ & $1.65\pm0.20$ \\
\hline
$\chic{0}\to \kp\km\ks\ks$ & $1.40\pm0.50$ \\
$\chic{1}\to \kp\km\ks\ks$ & $<0.4$ \\
$\chic{2}\to \kp\km\ks\ks$ & $<0.4$ \\
\hline \hline
\end{tabular}
%\caption{Branching fractions of $\chi_{cJ}$ into $2(\kp\km)$ and $\kp\km\ks\ks$ \cite{pdg}.}
\label{tab:bf}
\end{center}
\end{table}

\section{BESIII DETECTOR AND MONTE CARLO SIMULATION}
\label{sec:BES}
The BESIII detector is operated at the Beijing Electron Positron Collider~II~(BEPCII), which has reached a peak luminosity of $1.0\times10^{33}~\text{cm}^{-2}\text{s}^{-1}$ at a center-of-mass energy of $\sqrt s=3.773$~GeV. The detector has a geometrical acceptance of 93\% of the solid angle and is composed of four main components. A helium-gas based main drift chamber (MDC) is used to track charged particles. The single wire resolution is better than 130~$\mu$m, which, together with a magnetic field of 1~T, leads to a momentum resolution of 0.5\% at 1~GeV/$c$. The energy loss per path length $dE/dx$ is measured with a resolution of 6\%. The MDC is surrounded by a time-of-flight system built from plastic scintillators. It provides a $2\sigma ~K/\pi$ separation up to 1~GeV/$c$ momentum with a time resolution of 80 (110) ps for the barrel (end-caps). Particle energies are measured in the CsI(Tl) electro-magnetic calorimeter (EMC), which achieves an energy resolution for electrons of 2.5\% (5\%) at 1~GeV/$c$ momentum and a position resolution of 6~mm (9~mm) for the barrel (end-caps). Outside of the magnet coil, a muon counter composed of resistive plate chambers provides a spatial resolution of better than 2~cm. A more detailed description of the detector can be found in Ref.~\cite{NIM10}.

A \textsc{GEANT4}~\cite{Geant4} based Monte Carlo (MC) simulation package is used to optimize the event selections and estimate the signal efficiency and the background level. The event generator \textsc{KKMC}~\cite{Jadach01} simulates the electron-positron annihilation and the production of the $\psi$ resonances. Particle decays are generated by \textsc{EVTGEN}~\cite{Lange01} for the known decay modes with branching fractions from the Particle Data Group (PDG)~\cite{pdg} and \textsc{Lundcharm}~\cite{Lundcharm00} for the unknown ones. An inclusive MC sample containing $506\times10^6$ generic $\psi(3686)$ decays is used to study background.
The signal MC samples of the $\chi_{cJ}$ decays, generated according to a phase space model, are used to determine efficiencies.

\section{EVENT SELECTION}
\label{sec:selection}
\begin{figure}[b]
  %\centering
  \includegraphics[width=\columnwidth]{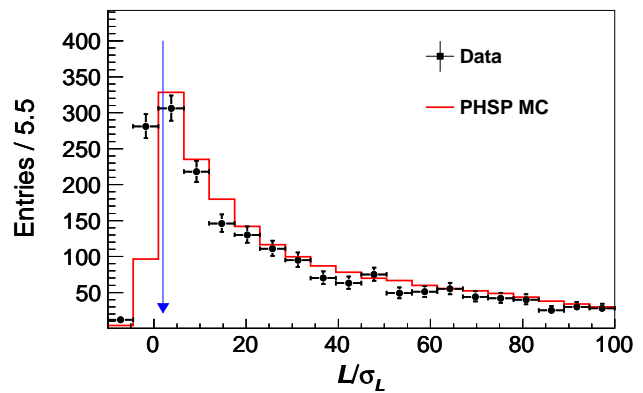}
  \caption{The distribution of $L/\sigma_{L}$ for all $K_{S}^{0}$ candidates. The arrow indicates the selection criteria, where the histogram is from the MC sample and the dots with error bars are from data.}
  \label{fig:sigma}
\end{figure}

\begin{figure}[t]
  \centering
  \includegraphics[width=\columnwidth]{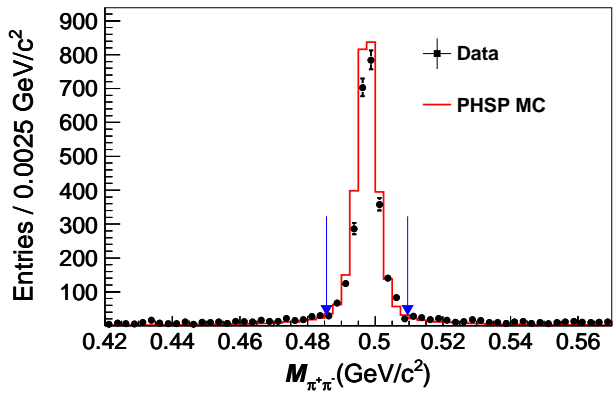}
  \caption{The $M_{\pi^+\pi^-}$ distribution for all $K_{S}^{0}$ candidates. The arrows indicate the mass window of the $\ks$ signal, where the histogram is from the MC sample and the dots with error bars are from data.}
  \label{fig:ks}
\end{figure}

\begin{figure}[b]
  \centering
  \includegraphics[width=\columnwidth]{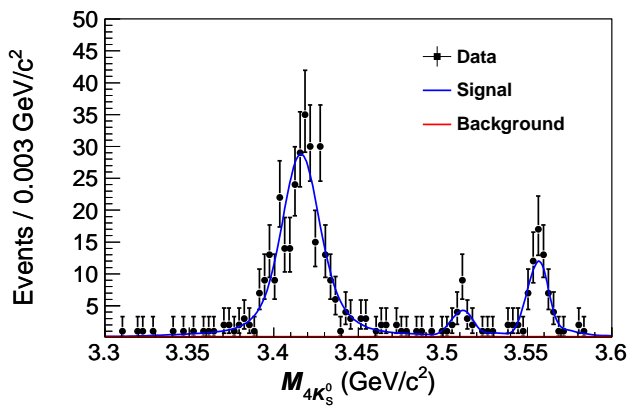}
  \caption{Fit to the $M_{4\ks}$ distribution of the candidate events of $\psi(3686)\to\chi_{cJ}$, $\chi_{cJ}\to 4\ks$. The points with error bars are data, the blue curve is the overall fit, and the red curve is the fitted background. }
  \label{fig:4ks}
\end{figure}

We reconstruct events from the decay chain of the charmonium
transitions $\psi(3686)\to\gamma\chic{J}$ followed by the hadronic
decays $\chic{J}\to4\ks$ and $\ks\to\pi^+\pi^-$. A photon candidate is defined as a
shower detected within the EMC exceeding an energy deposit of 25\mev in
the barrel region (covering the region $\rm |\cos\theta|<0.8$, where $\theta$ is the polar angle with respect to the positron beam direction) or of 50\mev in the end-caps ($0.86<\rm |\cos\theta|<0.92$). To suppress the electronics noise and beam background, the clusters are required to start within 700 ns after the event start time and fall outside a cone angle of $10^\circ$ around the nearest extrapolated charged track. All charged tracks are required to originate from the interaction region defined as $|V_{z}|<20$ cm and $|\rm \cos\theta|<0.93$, where $V_z$ denotes the distance of the closest approach of the reconstructed track to the interaction point (IP) in the $z$ direction.
Candidate events must have eight charged tracks with zero net charge and at least one good photon. The \ks candidates are reconstructed using vertex fits by looping over all oppositely charged track pairs in an event (assuming the tracks to be $\pi^{\pm}$ without particle identification). To suppress the $\pi^+\pi^-$ combinatorial background, the reconstructed decay lengths~($L$) of the \ks candidates are required to be more than twice their standard deviations~($\sigma_L$). The distribution of $L/\sigma_{L}$ for all $K_{S}^{0}$ candidates is shown in Fig.~\ref{fig:sigma}.
The invariant mass of $\pi^+\pi^-$ ($M_{\pi^+\pi^-}$) must be within the \ks signal region, defined as 12 MeV/$c^2$ around the \ks nominal mass~\cite{pdg}. The $M_{\pi^+\pi^-}$ distribution for all $K_{S}^{0}$ candidates is shown in Fig.~\ref{fig:ks}. To further suppress combinatorial background, a four-momentum conservation constraint (4C) is applied to the events. The $\chi_{\rm 4C}^{2}$ of the kinematic fit is required to be less than 200. The spectrum of the invariant mass of the $4K^0_S$~($M_{4K^0_S}$) of the accepted candidate events is shown in Fig.~\ref{fig:4ks}. Clear \chic{0}, \chic{1} and \chic{2} signals are observed.

%{\color{blue}The inclusive MC sample is used to investigate the potential peaking background. two events are found to survive by applying the same event selection as the real data. The details are discussed in Sec.~\ref{sec:background}}

The $M_{4K^0_S}$ distribution is fitted using an unbinned maximum likelihood fit. In the fit, each \chic{J} signal is
described with the MC simulation shape convolved with a Gaussian function with free parameters.
Since the background level is very low, as discussed in Sec.~\ref{sec:background}, the background shape is assumed to be flat. The signal yields of $\chi_{c0}$, $\chi_{c1}$ and $\chi_{c2}$ are fitted to be $319.4\pm19.0$, $21.6\pm5.2$ and $68.0\pm8.7$, respectively. The statistical significances are estimated to be 26.5$\sigma$, $5.9\sigma$ and $11.4\sigma$ for \chic{0}, \chic{1} and \chic{2} individually, which are determined by comparing the fit likelihood
values with and without each $\chi_{cJ}$ signal separately.

%{\color{red}\sout{We further examine the possible substructures in the $\chi_{cJ}\to4\ks$ decays which may affect the detection efficiencies. Fig.~\ref{fig:2ks3ks} show the distributions of invariant masses of 2$K_{S}^{0}$~($\rm M_{2K_{S}^{0}}$) and 3$K_{S}^{0}$~($\rm M_{3K_{S}^{0}}$). In the $\rm M_{2K_{S}^{0}}$ distribution, a bump around 1.5 GeV/$c^2$ is found. In $\rm M_{3K_{S}^{0}}$ distributions, no obvious structures are found. The systematic uncertainties due to the MC model are introduced to estimate the effect of possible substructures on the detection efficiencies. The details are discussed in Sec.~\ref{sec:systematics}.}}

\section{BACKGROUND STUDIES}
\label{sec:background}
The continuum data taken at $\sqrt s =$3.65 GeV, corresponding to an integrated luminosity of 44.45~pb$^{-1}$~\cite{lum}, are used to estimate the QED background. No events within this sample satisfy the same selection criteria applied to the main data sample.
In addition,
%{\color{red}\sout{the inclusive MC sample generated at 3.686 GeV is used to investigate the potential peaking background.}}
the inclusive MC sample is used to study all potential backgrounds from $\psi(3686)$ decays.
Two background events are found to be from $\psi(3686)\to \bar K^* K^0_S f^\prime_2$ and $\bar K^* K^0_S f_0(1710)$. Further studies with large exclusive MC samples show that the two background sources make up a uniform distribution around the $\chi_{cJ}$ signal regions. So, all peaking background components are negligible in this analysis.

%We further examine the substructures in the $\chi_{cJ}\to4\ks$ decays which may affect the detection efficiencies. In the spectrum of the invariant mass of $K_{S}^{0}K_{S}^{0}$, a peaking-like structure around 1.5 GeV is found. The effect due to this possible sub-resonance will be investigated in Sec.~\ref{sec:systematics}. In the spectrum of the invariant mass of $K_{S}^{0}K_{S}^{0}K_{S}^{0}$, no peaking-like structure is found.

\section{BRANCHING FRACTIONS}
\label{sec:mc}
%A detailed MC simulation of the \besiii detector based
%on {\sc geant4} \cite{geant4} is used to determine efficiencies, signal shapes,
%background contributions and systematic uncertainty. The production of the \psiprime resonance is
%simulated using the {\sc kkmc} event generator \cite{kkmc}. Decays of
%the \psiprime and subsequent particles in the event are modeled by
%{\sc evtgen} \cite{evtgen}. Simulated events pass the same reconstruction
%algorithms and selection criteria as data.

\begin{figure}[b]
  \centering
  \includegraphics[width=\columnwidth]{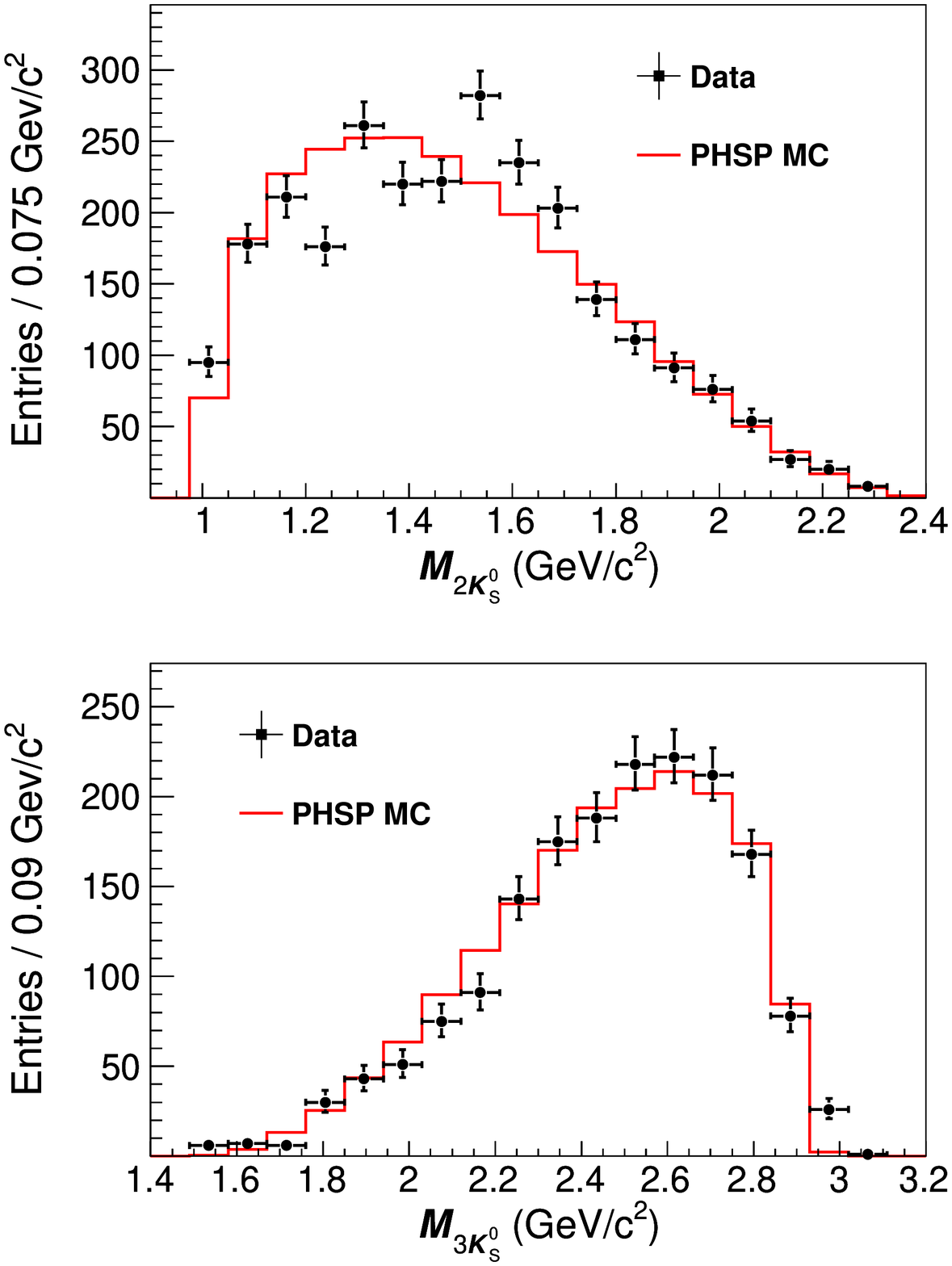}
  \caption{The $M_{2K_{S}^{0}}$ and $M_{3K_{S}^{0}}$ distributions for all 2$K_{S}^{0}$ and 3$K_{S}^{0}$ combinations, where the histogram is from the MC sample and the dots with error bars are from data.}
  \label{fig:2ks3ks}
\end{figure}

\begin{figure}[b]
  \centering
  \includegraphics[width=\columnwidth]{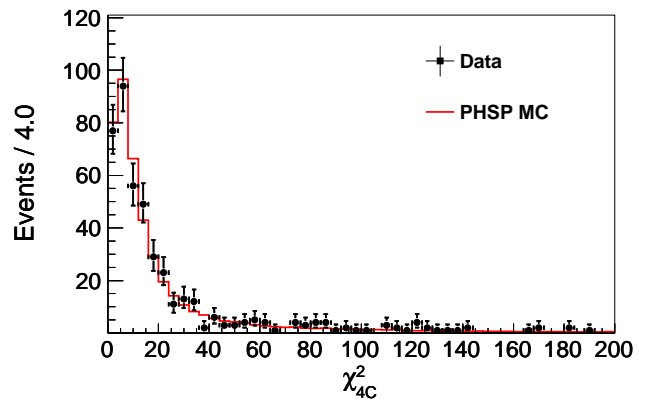}
  \caption{The $\chi_{\rm 4C}^{2}$ distribution after corrections, where the histogram is from the MC sample and the dots with error bars are from data.}
  \label{fig:compare}
\end{figure}

For each decay of
$\psi(3686)\to\gamma\chic{J}$, $\chic{J}\to4\ks$, $K_{S}^{0}\to\pi^+\pi^-$, $5\times10^5$ signal MC events are generated
using a $1+\lambda\cos^2\theta$ distribution, where $\theta$ is the angle between the direction of the radiative photon and the
beam, and $\lambda=1,-1/3,1/13$ for $J=0,1,2$ in accordance with
expectations for electric dipole transitions. Since no obvious substructures are found in the $M_{2K^0_S}$ and $M_{3K^0_S}$
distributions of the accepted $\chi_{cJ}\to 4\ks$ candidate events, as shown in Fig.~\ref{fig:2ks3ks}, the \chic{J} decay
products are generated using phase space~(PHSP). Intrinsic width and mass values as given in Ref.~\cite{pdg} are used for the
\chic{J} states in the simulation. To reduce the difference of the distributions of $\chi^2$ of the 4C kinematic fit ($\chi_{4\rm C}^{2}$) between data and MC simulation, we correct the track helix parameters of MC simulation in the 4C kinematic fit. The $\chi_{4\rm C}^{2}$ distribution after corrections is shown in Fig.~\ref{fig:compare}, in which the consistency between data and MC simulation is reasonable. The obtained corrected efficiencies for
$\chi_{cJ}\to4K^0_S$ are $(5.51 \pm 0.03)\%$, $(6.19 \pm 0.04)\%$ and $(6.08 \pm 0.04)\%$, respectively, including detector acceptance as well as reconstruction
and selection efficiencies.

%The continuum data taken at $\sqrt s= 3.65$ GeV is used to estimate the QED backgrounds. No events are found to survive the event selection.
%
%In addition, we use inclusive MC samples to investigate backgrounds. Two background events are found to be from $\psi^\prime\to \bar K^* K^0_S f^\prime_2$ and $\bar K^* K^0_S f_0(1710)$. Further MC studies show that the two dominant background sources do not form peak around the $\chi_{cJ}$ signal regions.

%We further examine the substructures in the $\chi_{cJ}\to4\ks$ decays which may affect the detection efficiencies. In the spectrum of the invariant mass of $K_{S}^{0}K_{S}^{0}$, a peaking-like structure around 1.5 GeV is found. The effect due to this possible sub-resonance will be investigated in Sec.~\ref{sec:systematics}. In the spectrum of the invariant mass of $K_{S}^{0}K_{S}^{0}K_{S}^{0}$, no peaking-like structure is found.

%\begin{table}[htbp]
%   \centering
%   \begin{tabular}{l|c|c|c}
%   \hline
%   \hline
%        \text{No}      & \text{Decay chain }& \text{Final states}& \text{nEvt}\\
%        \hline
%        \text{1} & \text{$\psi^{'}\to \bar{K^{*}}K_{S}^{0}f_{2}^{'}$}&\text{$\psi^{'}\to \pi^{+}\pi^{+}\pi^{+}\pi^{+}\pi^{0}\pi^{-}\pi^{-}\pi^{-}\pi^{-}$}&\text{$1$}\\
%        \hline
%        \text{2} & \text{$\psi^{'}\to \bar{K^{*}}K_{S}^{0}f_{0}(1710)$}&\text{$\psi^{'}\to \pi^{+}\pi^{+}\pi^{+}\pi^{+}\pi^{0}\pi^{-}\pi^{-}\pi^{-}\pi^{-}$}&\text{$1$}\\
%        \hline
%        \hline
%
%    \end{tabular}
%    \caption{Topologies of the backgrounds in $\chi_{cJ}\to 4K^0_S$ from inclusive MC sample.}
%    \label{tab:topo}
%\end{table}
The signal yields $N_{\rm obs}^{J}$ are obtained by fitting to the $M_{4K^0_S}$ distribution. The branching fraction is calculated with

\begin{small}
\begin{equation}
\mathcal{B}_{\chi_{cJ}\to 4K_{S}^{0}}=\frac{N_{\rm obs}^{ J}}{N_{\psi(3686)}\cdot\mathcal{B}_{\psi(3686)\to\gamma\chi_{cJ}}\cdot\mathcal{B}^{4}_{K_{S}^{0}\to \pi^{+}\pi^{-}}\cdot\epsilon},
\end{equation}
\end{small}where $\epsilon$ is the efficiency, $N_{\psi(3686)}$ is the number of
$\psi(3686)$ events,
$\mathcal{B}_{\psi(3686)\to\gamma\chic{J}}$ and
$\mathcal{B}_{\ks\to\pip\pim}$ are the branching fractions of the PDG fit of
$\psi(3686)\to\gamma\chi_{cJ}$ decays and $K_{S}^{0}\to\pi^+\pi^-$ decay~\cite{pdg}.

\section{SYSTEMATIC UNCERTAINTIES}
\label{sec:systematics}
The systematic uncertainties in the measurements of $\mathcal{B}_{\chi_{cJ}\to 4K^0_S}$ originate
from several sources, as summarized in Table~\ref{tab:systematics}. They are estimated and described below.\\
\begin{table}[t]
  %\begin{center}
  \caption{Summary of the systematic uncertainties~(\%). }
  \begin{tabular} {lrrr}
  \hline\hline
  Source                             &  $\chi_{c0}$ &  $\chi_{c1}$ &  $\chi_{c2}$ \\
  \hline
  Number of $\psi(3686)$ events                     & 0.6  & 0.6  & 0.6   \\
  $\gamma$ detection                                & 1.0  & 1.0  & 1.0   \\
  \ks reconstruction                                & 6.0  & 6.0  & 6.0   \\
  MC model                                          & 0.4  & 0.2  & 0.2   \\
  4C kinematic fit                                  & 1.0  & 1.0  & 1.0   \\
  Angular distribution                              & 0.7  & 0.5  & 0.7   \\
  Fit range                                         & 0.6  & 1.5  & 0.9   \\
  Signal shape                                      & 0.4  & 2.8  & 1.7   \\
  MC statistics                                     & 0.6  & 0.5  & 0.6   \\
  Quoted branching fractions                        & 2.0  & 2.5  & 2.1   \\
  %$\mathcal{B}(\ks\to\pip\pim)$                     & 0.3  & 0.3  & 0.3   \\
  \hline
  Total                                             & 6.6  & 7.4  & 6.9   \\
  \hline \hline
  \end{tabular}
  %\caption{Summary of the systematic uncertainties. }
  \label{tab:systematics}
  %\end{center}
\end{table}
The number of $\psi(3686)$ events has been measured to be $N_{\psi(3686)}=(448.1\pm2.9)\times10^6$ with the inclusive
hadronic data sample, as described in Ref.~\cite{ref::psip-num-inc}. The uncertainty of the total number is 0.6\%.

The systematic uncertainty due to the photon detection is assumed to be 1.0\% per photon with the control sample $J/\psi\to\rho^0\pi^0$~\cite{ref::gamma-recon}.

The systematic uncertainty associated with \ks reconstruction is determined to be 1.5\% per
\ks with the control samples of $J/\psi\to K^{*\pm}(892)K^{\mp}$, $K^{*\pm}(892)\to K^0_S\pi^{\pm}$ and $J/\psi\to\phi K_{S}^{0}K^{\mp}\pi^{\pm}$ in Ref.~\cite{ref::ks}. %in which the uncertainty due to $K^0_S$ reconstruction has been determined to be $(1.01\pm0.53)\%$.

%{\color{red}\sout{We further examine the possible substructures in the $\chi_{cJ}\to4\ks$ decays which may affect the detection efficiencies. In the spectrum of the invariant mass of $K_{S}^{0}K_{S}^{0}$, a peaking-like structure around 1.5 GeV is found. In the spectrum of the invariant mass of $K_{S}^{0}K_{S}^{0}K_{S}^{0}$, no peaking-like structure is found.}}
 To estimate the systematic uncertainties of the MC model for the $\chi_{cJ}\to4K_{S}^{0}$ decay, we compare our nominal efficiency with that determined from the signal MC events after mixing
some possible sub-resonant decays, including $\chi_{cJ}\to f_{0}(1500)f_{0}(1500)$, $\chi_{cJ}\to K_{S}^{0}K_{S}^{0}f_{0}(1500)$, $\chi_{cJ}\to K_{S}^{0}K_{S}^{0}f_{2}^{'}(1525)$, $\chi_{cJ}\to f_{0}(1500)f_{2}^{'}(1525)$, $\chi_{cJ}\to f_{0}(1500)f_{0}(1710)$, $\chi_{cJ}\to f_{0}(1500)f_{2}(1565)$
and $\chi_{cJ}\to f_{2}^{'}(1525)f_{2}(1565)$. The systematic uncertainties are estimated as the relative changes of efficiencies, which are 0.4\%, 0.2\% and 0.2\% for \chic{0}, \chic{1} and \chic{2} decays, respectively.

We correct the track helix parameters for MC simulation in the 4C kinematic fit. The change in detection efficiency is not more than 1.0\% when varying the correction factors within one standard deviation around the nominal value. We therefore assume 1.0\% as the systematic uncertainty of the 4C kinematic fit.

To estimate the systematic uncertainties in the angular distribution, we use a reweighting method. New signal MC events are obtained by reweighting the angular distribution of the $K^0_S$ in the signal MC events to data. The changes to the detection efficiencies are taken as the systematic uncertainties, which are 0.7\%, 0.5\% and 0.7\% for $\chi_{c0}$, $\chi_{c1}$ and $\chi_{c2}$ decays, respectively. % (see Table~\ref{tab:systematics}).

The systematic uncertainties due to the fit range are estimated by a series of fits with alternative intervals.% {\color{red}\sout{of (3.30,\,3.59), (3.30,\,3.61), (3.30,\,3.62), (3.28,\,3.60), (3.29,\,3.60), (3.31,\,3.60) and (3.32,\,3.60) GeV/$c^2$}}.
~The standard deviations of the resulting branching fractions are assigned as the
systematic uncertainties, which are 0.6\%, 1.5\% and 0.9\% for $\chi_{c0}$, $\chi_{c1}$ and $\chi_{c2}$ decays, respectively.  % (see Table~\ref{tab:systematics}).

To estimate the systematic uncertainties due to the signal shape, we use alternative signal shapes, a Breit Wigner function smeared with a double Gaussian and a MC shape (including $\rm E^3$ dependence) convolved with a Gaussian function, to describe each \chic{J} signal. The maximum deviations of the resulting branching fractions are assigned as the relevant systematic uncertainties, which are 0.4\%, 2.8\% and 1.7\% for $\chi_{c0}$, $\chi_{c1}$ and $\chi_{c2}$ decays, respectively. % (see Table~\ref{tab:systematics}).

The systematic uncertainties due to the statistics of the MC samples %{\color{red}\sout{is calculated by}}
%\begin{equation}
%\sqrt{\frac{(1-\epsilon)/\epsilon}{N}}{\color{blue},}
%\end{equation}
%{\color{red}\sout{where $\epsilon$ is detection efficiency and ${\color{blue}N}$ is total number of MC. They are}}
are 0.6\%, 0.5\%, and 0.6\% for $\chi_{c0}$, $\chi_{c1}$ and $\chi_{c2}$ decays, respectively.

%The uncertainties due to the quoted branching fractions $\psiprime\to\gamma\chic{J}$ are 2.7\% for
%\chic{0}, 3.2\% for \chic{1}, and 3.4\% for \chic{2}.
%The uncertainties due to the quoted branching fractions $\ks\to\pip\pim$ are 0.3\% for \chic{0}, \chic{1} and \chic{2}.
The systematic uncertainties from the branching fractions of $\psi(3686)\to\gamma\chic{J}$ and $\ks\to\pip\pim$ decays quoted from the PDG~\cite{pdg} are 2.0\%, 2.5\% and 2.1\% for $\chi_{c0}$, $\chi_{c1}$ and $\chi_{c2}$ decays and 0.07\% for $K_S^0$, respectively.

We assume that all systematic uncertainties are independent and
add them in quadrature to obtain the total systematic uncertainty for each decay.

\section{Conclusion}
By analyzing $(448.1\pm2.9)\times10^6$ $\psi(3686)$ events with the BESIII detector,
the product branching fractions are determined to be $\mathcal{B}_{\psi(3686)\to\gamma\chi_{c0}}\times\mathcal{B}_{\chic{0}\to4\ks}=$$(0.564\pm0.033\pm0.037)$$\times10^{-4}$, $\mathcal{B}_{\psi(3686)\to\gamma\chi_{c1}}\times\mathcal{B}_{\chic{1}\to4\ks}=$$(0.034\pm0.009\pm0.003)$$\times10^{-4}$ and $\mathcal{B}_{\psi(3686)\to\gamma\chi_{c2}}\times\mathcal{B}_{\chic{2}\to4\ks}=$$(0.108\pm0.015\pm0.008)$$\times10^{-4}$, where the uncertainties are statistical and systematic. We measure for the first time the branching fractions of
$\chic{J}\to4\ks$ decays to be $\mathcal{B}_{\chic{0}\to4\ks}=$$(5.76\pm0.34\pm0.38)$$\times10^{-4}$,
$\mathcal{B}_{\chic{1}\to4\ks}$=$(0.35\pm0.09\pm0.03)$$\times10^{-4}$, $\mathcal{B}_{\chic{2}\to4\ks}$=$(1.14\pm0.15\pm0.08)$$\times10^{-4}$, where the first and second
uncertainties are statistical and systematic, respectively. %{\color{blue}The product branching fractions are determined to be $\mathcal{B}_{\psi(3686)\to\gamma\chi_{c0}}\times\mathcal{B}_{\chic{0}\to4\ks}=$$(0.564\pm0.054)$$\times10^{-4}$, $\mathcal{B}_{\psi(3686)\to\gamma\chi_{c1}}\times\mathcal{B}_{\chic{1}\to4\ks}=$$(0.034\pm0.009)$$\times10^{-4}$ and $\mathcal{B}_{\psi(3686)\to\gamma\chi_{c2}}\times\mathcal{B}_{\chic{2}\to4\ks}=$$(0.108\pm0.017)$$\times10^{-4}$.} At present, there is still a large amount of unobserved decay width for the $\chi_{cJ}$ mesons~\cite{pdg}. Our measurements improve the existing knowledge of the \chic{J} states and benefit the understanding of isospin symmetry by combining with the information of $\chi_{cJ}\to2(K^+K^-)$ and $\chi_{cJ}\to K_{S}^{0}K_{S}^{0}K^+K^-$. %and improve our understanding of the dynamics mechanism of processes of multi-body decays.

\section{ACKNOWLEDGMENTS}
%The BESIII collaboration thanks the staff of BEPCII and the IHEP computing center for their strong support. This work is supported in part by National
%Key Basic Research Program of China under Contract No. 2015CB856700; National Natural Science Foundation of China (NSFC) under Contracts Nos. 11235011,
%11335008, 11425524, 11505034, 11575077, 11625523, 11635010, 11675200, 11775230; the Chinese Academy of Sciences (CAS) Large-Scale Scientific Facility Program; the CAS Center for Excellence in Particle Physics (CCEPP); Joint Large-Scale Scientific Facility Funds of
%the NSFC and CAS under Contracts Nos. U1332201, U1532257, U1532258, U1632109; CAS under Contracts Nos. KJCX2-YW-N29, KJCX2-YW-N45, QYZDJ-SSW-SLH003; 100 Talents Program of CAS; National 1000 Talents Program of China; INPAC and Shanghai Key Laboratory for Particle Physics and Cosmology; German Research Foundation DFG under Contracts Nos. Collaborative Research Center CRC 1044, FOR 2359; Istituto Nazionale di Fisica Nucleare, Italy; Koninklijke Nederlandse Akademie van Wetenschappen (KNAW) under Contract No. 530-4CDP03; Ministry of Development of Turkey under Contract No. DPT2006K-120470;
%National Science and Technology fund; The Swedish Research Council; U. S. Department of Energy under
%Contracts Nos. DE-FG02-05ER41374, DE-SC-0010118,
%DE-SC-0010504, DE-SC-0012069; University of Groningen (RuG) and the Helmholtzzentrum fuer Schwerionenforschung GmbH (GSI), Darmstadt; WCU Program of
%National Research Foundation of Korea under Contract No. R32-2008-000-10155-0.

The BESIII collaboration thanks the staff of BEPCII and the IHEP computing center for their strong support. This work is supported in part by National Key Basic Research Program of China under Contract No. 2015CB856700; National Natural Science Foundation of China (NSFC) under Contracts Nos. 11335008, 11425524, 11625523, 11635010, 11735014; the Chinese Academy of Sciences (CAS) Large-Scale Scientific Facility Program; the CAS Center for Excellence in Particle Physics (CCEPP); Joint Large-Scale Scientific Facility Funds of the NSFC and CAS under Contracts Nos. U1532257, U1532258, U1732263; CAS Key Research Program of Frontier Sciences under Contracts Nos. QYZDJ-SSW-SLH003, QYZDJ-SSW-SLH040; 100 Talents Program of CAS; INPAC and Shanghai Key Laboratory for Particle Physics and Cosmology; German Research Foundation DFG under Contracts Nos. Collaborative Research Center CRC 1044, FOR 2359; Istituto Nazionale di Fisica Nucleare, Italy; Koninklijke Nederlandse Akademie van Wetenschappen (KNAW) under Contract No. 530-4CDP03; Ministry of Development of Turkey under Contract No. DPT2006K-120470; National Science and Technology fund; The Swedish Research Council; U. S. Department of Energy under Contracts Nos. DE-FG02-05ER41374, DE-SC-0010118, DE-SC-0010504, DE-SC-0012069; University of Groningen (RuG) and the Helmholtzzentrum fuer Schwerionenforschung GmbH (GSI), Darmstadt.


\begin{thebibliography}{99}
\bibitem{com} G.~T.~Bodwin, E.~Braaten and G.~P.~Lepage, \jprd{\bf 51}, 1125 (1995);
              H.~W.~Huang and K.~T.~Chao, \jprd{\bf 54}, 6850 (1996);
              A.~Petrelli, \plb{\bf 380}, 159 (1996);
              J.~Bolz, P.~Kroll and G.~A.~Schuler, Eur. Phys. J. {\bf C~2}, 705 (1998);
              S.~H.~M.~Wong, Eur. Phys. J. {\bf C~14}, 643 (2000).
\bibitem{ref::theroy1} R. G. Ping, B. S. Zou and H. C. Chiang, Eur.~Phys.~J.~{\bf A 23}, 129 (2005).
\bibitem{ref::theroy2} X. H. Liu and Q. Zhao, J. Phys. {\bf G 38}, 035007 (2011).
\bibitem{ref::theroy3} S. M. H. Wong, Eur. Phy. J. {\bf C 14}, 643 (2000).
\bibitem{pdg} M. Tanabashi {\it et al.} (Particle Data Group), Phys. Rev. {\bf D 98}, 020001 (2018).
\bibitem{ref::kkkk} S. Uehara {\it et al.}, Eur.~Phys.~J.~{\bf C 53}, 1 (2008).
\bibitem{ref::kkk0k0} M. Ablikim {\it et al.} (BESIII Collaboration), Phys. Lett. {\bf B 630}, 21 (2005).
\bibitem{ref::psip-num-inc} M. Ablikim {\it et al.} (BESIII Collaboration), Chin. Phys. {\bf C 42}, 023001 (2018).
\bibitem{NIM10} M.~Ablikim \textit{et al.} (BESIII Collaboration), Nucl. Instrum. Methods Phys. Res., Sect. {\bf A 614}, 345 (2010).
%\bibitem{pdg} C. Patrignani {\it et al.} (Particle Data Group), Chin. Phys. {\bf C 40}, 100001 (2016).
%\bibitem{NIM10} M.~Ablikim \textit{et al.} (BESIII Collaboration), Nucl. Instrum. Methods Phys. Res., Sect. {\bf A 614}, 345 (2010).
\bibitem{Geant4} S.~Agostinelli \textit{et al.} ({\sc geant4} Collaboration), Nucl. Instrum. Methods Phys. Res., Sect. {\bf A 506}, 250 (2003).
\bibitem{Jadach01} S.~Jadach, B.~F.~L.~Ward~and Z.~Was, Phys. Rev. {\bf D~63}, 113009 (2001).
\bibitem{Lange01} D.~J.~Lange, Nucl. Instrum. Methods Phys. Res., Sect. {\bf A 462}, 152 (2001);  R.~G.~Ping, Chin. Phys. {\bf C~32}, 599 (2008).
\bibitem{Lundcharm00} J.~C.~Chen, G.~S.~Huang, X.~R.~Qi, D.~H.~Zhang and Y.~S.~Zhu, Phys. Rev. {\bf D~62}, 034003 (2000).
\bibitem{lum} M. Ablikim {\it et al.} (BESIII Collaboration), Chin. Phys. {\bf C 37}, 123001 (2013)
%\bibitem{geant4} S.~Agostinelli \etal ({\sc geant4} Collaboration), Nucl. Instrum. Meth. {\bf A~506}, 250 (2003).

%\bibitem{kkmc} S.~Jadach, B.F.L. Ward and Z.~Was, Comput. Phys. Commun. {\bf 130}, 260 (2000);
%               S.~Jadach, B.F.L. Ward and Z.~Was \jprd{\bf 63}, 113009 (2001).
%\bibitem{evtgen} D.J.~Lange \etal, Nucl. Instrum. Meth. {\bf A~462}, 1 (2001).
%\bibitem{ref::psip-num-inc} M. Ablikim {\it et al.} (BESIII Collaboration), arXiv:1709.03653 [hep-ex], accepted by Chin. Phys. {\bf C}.
%\bibitem{pdg2018} {\color{red}\sout{M. Tanabashi {\it et al.} (Particle Data Group), Phys. Rev. {\bf D 98}, 020001 (2018).}}
\bibitem{ref::gamma-recon} M. Ablikim {\it et al.} (BESIII Collaboration), Phys. Rev. {\bf D 86}, 052011 (2012). %BESIII analysis memo: BAM-00038, {\it Measurement of $\chi_{cj}$ decaying into $p\bar{n}\pi^-+c.c.$ and $p\bar{n}\pi^-\pi^0+c.c.$ final states}.
\bibitem{ref::ks} M. Ablikim {\it et al.} (BESIII Collaboration), Phys. Rev. {\bf D 92}, 112008 (2015).


\end{thebibliography}
\end{document}